\documentclass[conference]{IEEEtran}
\IEEEoverridecommandlockouts
\usepackage{cite}
\usepackage{amsmath,amssymb,amsfonts}
\usepackage{algorithmic}
\usepackage{graphicx}
\usepackage{textcomp}
\usepackage{xcolor}
\def\BibTeX{{\rm B\kern-.05em{\sc i\kern-.025em b}\kern-.08em
    T\kern-.1667em\lower.7ex\hbox{E}\kern-.125emX}}
\begin{document}

\title{A Survey on Super Resolution for video Enhancement Using GAN\\
}

\author{\IEEEauthorblockN{1\textsuperscript{st} Ankush Maity}
\IEEEauthorblockA{\textit{Dept. of Computer Engineering} \\
\textit{Army Institute of Technology}\\
Pune,India \\
ankushmaity\_20512@aitpune.edu.in}
\and
\IEEEauthorblockN{2\textsuperscript{nd} Sourabh Kumar Lenka}
\IEEEauthorblockA{\textit{Dept. of Computer Engineering} \\
\textit{Army Institute of Technology}\\
Pune,India \\
sourabhlenka\_20091@aitpune.edu.in}
\and
\IEEEauthorblockN{3\textsuperscript{rd} Roshan Pious}
\IEEEauthorblockA{\textit{Dept. of Computer Engineering} \\
\textit{Army Institute of Technology}\\
Pune,India \\
roshanpious\_20155@aitpune.edu.in}
\and
\IEEEauthorblockN{4\textsuperscript{th} Vishal Choudhary}
\IEEEauthorblockA{\textit{Dept. of Computer Engineering} \\
\textit{Army Institute of Technology}\\
Pune,India \\
vishalchoudhary\_20856@aitpune.edu.in}
\and
\IEEEauthorblockN{5\textsuperscript{th} Prof. Sharayu Lokhande}
\IEEEauthorblockA{\textit{Dept. of Computer Engineering} \\
\textit{Army Institute of Technology}\\
Pune,India \\
slokhande@aitpune.edu.in}
}

\maketitle

\begin{abstract}
This compilation of various research paper highlights provides a comprehensive overview of recent developments in super-resolution image and video using deep learning algorithms such as Generative Adversarial Networks. The studies covered in these summaries provide fresh techniques to addressing the issues of improving image and video quality, such as recursive learning for video super-resolution, novel loss functions, frame-rate enhancement, and attention model integration. These approaches are frequently evaluated using criteria such as PSNR, SSIM, and perceptual indices. These advancements, which aim to increase the visual clarity and quality of low-resolution video, have tremendous potential in a variety of sectors ranging from surveillance technology to medical imaging. In addition, this collection delves into the wider field of Generative Adversarial Networks, exploring their principles, training approaches, and applications across a broad range of domains, while also emphasizing the challenges and opportunities for future research in this rapidly advancing and changing field of artificial intelligence.
\end{abstract}


\section{Introduction}
Improving picture and video quality has long been a goal in the fields of computer vision and image processing. Deep learning techniques, along with the unique use of Generative Adversarial Networks (GANs), have resulted in significant improvements in the field of image and video super-resolution in recent years. The results of these discoveries have not only improved the visual acuity of the low-resolution material, but have also led to the development of new areas of investigation, such as surveillance, medical imaging and remote sensing.

This compilation of literature review papers aims to provide a comprehensive review of the most recent developments in image and video super-resolution, with a special emphasis on the integration of deep learning approaches and GANs. The studies reviewed here introduce unique ideas, each tackling specific issues and pushing the frontiers of what is possible in the hunt for higher-resolution images.

The use of GANs as a framework for creating images and videos with high perceptual quality and realism is a significant theme studied in these publications. These methods use GANs to create information that not only closely resembles the original high-resolution images, but also has the depth of detail and visual quality that can be critical in a variety of applications.

The literature reviews cover a broad array of subject areas related to image and video super resolution, including single image super resolution and video super resolution. They go into the complexities of loss functions, recursive learning, edge enhancement, and frame rate enhancement, illuminating how these techniques contribute to overall image and video quality improvement.

The assessment of these approaches is a fundamental part of the literature discussed in this paper, utilizing metrics such as the Peak Signal-to-Number (PSNR), the Structural Similarity Indicator (SSIM), and the Perceptual Indices (PIs) to evaluate the effectiveness of the proposed approaches. Additionally, these articles address the challenges associated with deep learning model training, the need for enormous quantities of labelled data, and the intricacies of degradation processes encountered in real-world conditions.

We will investigate the new approaches, the potential impact on diverse sectors, and the future directions that this quickly evolving field has as we go into each of these literature studies. These studies demonstrate the transformative impact of deep learning and GANs in the pursuit of improved visual quality in photos and videos, ranging from the enhancement of CCTV footage to the restoration of medical images.

\section{Literature Review}
The author of [1] examines the field of image super-resolution in computer vision, which aims to improve image quality by increasing spatial dimensions. The task of achieving high perceptual quality is difficult and involves algorithms reconstructing high-quality images from low-resolution counterparts. In recent years, deep learning has been used to enhance the texture of low-resolution images. However, obtaining corresponding image pairs for training purposes is challenging, often resulting in the generation of synthetic low-resolution images. In order to address the challenge of achieving super-resolution results from low-resolution images with unknown degradation, researchers propose a novel two-stage GAN system, with two architectural variations. This approach demonstrates the potential of learning to down-scale images in a low-level supervised environment, providing a powerful technology to improve super-resolution results. In conclusion, image super-resolution is a complex task, but this new GAN-based approach exhibits potential in overcoming the limitations of synthetic data and enhancing real-world image quality.The future work seeks to address the inherent issues associated with this ill-conceived problem, where images of low resolution can have multiple interpretations in high-resolution, resulting in subjectivity in the quality of perception. Scientists are currently investigating the possibility of learning to reduce noise in low-resolution data to enhance super-resolution output. This involves training models to denoise image degradations, potentially enhancing texture and detail reconstruction. In practical applications, super-resolution holds promise in surveillance technology by improving the quality of recorded footage, facilitating better monitoring and decision-making without the need for expensive equipment. Super-resolution in computer vision provides a cost-effective approach to enhance image quality, preserving historical information and delivering high-fidelity content. However, existing techniques struggle with generalization, resulting in the presence of blurry artifacts. Researchers are investigating methods to learn real-world degradation for super-resolution, with encouraging outcomes in reducing the reliance on supervised image pairs. Nevertheless, addressing low-resolution noise may necessitate the utilization of dense networks for more efficient feature extraction from low-resolution images.

In [2], The authors introduce an advanced super-resolution reconstruction algorithm that is based on the idea of a  GAN. The goal of this algorithm is to increase the resolution of low resolution images by creating high-resolution images with enhanced details and visual effects. What sets this algorithm apart is the refined network model and loss function employed, both of which have undergone significant improvements and optimization. The incorporation of an auxiliary VGG-19 network aids in the extraction of image features, while an extended convolution expands the receptive field, culminating in more impressive reconstruction outcomes. The algorithm is honed through training using the DIV2k dataset as the training set and set5, set14, and bsd100 datasets as the testing set. The authors conduct an experimental analysis to demonstrate the feasibility of the proposed method and demonstrate the perceptible improvement in image quality in comparison to existing conventional models. The article provides an in-depth analysis of the complex structure of the proposed generative net and discriminant network models, elucidating their intrinsic mechanisms. During the training process, low-resolution images are fed into the generation network to yield super-resolved images, which are then subjected to a VGG-19 network for feature map extraction. The discrimination results are generated by comparison of the feature maps of high resolution images and the resulting loss. The total loss, comprised of multiple loss functions, is employed to train and update the network parameters. The experiments are conducted in a variety of environments and the results are evaluated using a variety of parameters, including the PSNR, SSIM, and MOS. The outcome of these evaluations demonstrates that the proposed method surpasses traditional convolutional neural networks, offering superior visual effects and higher MOS scores. The authors also highlight the necessity for more robust evaluation indicators to accurately assess the quality of image reconstruction. In conclusion, the proposed enhanced super-resolution reconstruction algorithm, based on a generative adversarial network, exhibits promising potential in achieving high-quality and aesthetically pleasing image super resolution.

The introduction of Paper [3] presents an innovative technique for enhancing image resolution known as MR-SRGAN (Multi-branch receptive field dense block improved super-resolution generative adversarial network). This method aims to tackle the challenges of accurately extracting intricate texture features and achieving network convergence during the training phase. To overcome these obstacles, the MR-SRGAN algorithm incorporates the use of the new MBRS residual block and MRB module for extracting detailed texture features from images, while also improving training convergence by adjusting the loss function. In order to evaluate its performance, the algorithm is compared to the SRGAN algorithm through super-resolution experiments conducted on the Set5 and Set14 datasets. The results demonstrate that MR-SRGAN outperforms SRGAN by achieving a higher PSNR ratio, thus demonstrating the superiority of the reconstruction quality. The capability to improve the resolution of imagery is essential for a variety of applications, including satellite reconnaissance, medical imaging, and industrial control, as it allows for clear visualization of information. Traditional methods that rely on interpolation often result in blurriness and visual fragmentation, while reconstruction-based approaches heavily rely on prior knowledge, making them susceptible to quality degradation when dealing with high magnification or limited input images. On the other hand, depth learning-based methods have exhibited better performance, with SRGAN being an advanced algorithm in this domain. Nonetheless, there is still room for improvement when it comes to accurately extracting texture features from images and achieving network convergence.The MR-SRGAN algorithm introduces the novel MBRS residual block and MRB module to enhance feature extraction and effectively address the convergence issue identified in the SRGAN algorithm. Through experimental evaluations, it has been observed that MR-SRGAN produces images with more intricate texture details and better edge information compared to other algorithms. Objectively, the algorithm achieves higher PSNR and SSIM values, indicating superior reconstruction quality. All in all, the MR-SRGAN algorithm demonstrates promising results in enhancing image super-resolution by effectively addressing the limitations present in existing methods. Future research endeavors could focus on further refining the model to expedite training and improve accuracy.

In [4] the author explores the utilization of a GAN (Generative Adversarial Network) to enhance the clarity of low-resolution images. The GAN network employs a discriminator to distinguish between genuine and generated images, and it possesses the ability to restore edges and amplify the overall visual excellence of the low-resolution images. The article introduces a suggested technique for Image Super-Resolution through the utilization of the GAN approach. The writers refer to past studies on GANs for Super-Resolution and emphasize the effectiveness of deep residual networks and advanced deep CNNs for enhancing images. They also address the issue of insufficient feature extraction and propose a GAN model that permits greater adaptability and acceptance. The suggested technique involves two facets: capturing images using a micro lens array (MLA) and enhancing the resolution using the recommended GAN technique. The captured images undergo processing, and their resolution is enhanced before being merged for the complete visualization system. A honeybee example is employed, and a specific area of interest is chosen to diminish noise. The article further elaborates on the training of the GAN model using a loss function, supervised learning, and the use of PSNR and other image quality indicators to assess the model's performance. The authors conducted experiments using various image inputs and evaluated the quality of the generated images.They discovered that the suggested model performs admirably based on metrics such as PSD, SSIM, and PSNR. Additionally, it generates top-notch, high-resolution images in a brief span of time. The article introduces a GAN-centered methodology for enhancing the resolution of low-quality images. The recommended technique demonstrates encouraging outcomes in enhancing image quality and producing high-resolution images.

Article [5] presents a yardstick for examining the ability of image and video super-resolution (SR) models to restore intricate details. The authors crafted a dataset encompassing intricate patterns that SR models often falter in restoring accurately. This methodology was used to evaluate 32 recent SR models and compare their ability to maintain the context of the scene. To verify the authenticity of the restored details, a comparative analysis of the restored details was conducted using crowd-sourced data and an objective evaluation metric was developed, referred to as ERQAV2.0. The findings unveiled that numerous SR methods concentrate on enhancing the authenticity of the resulting image but may inadvertently generate flawed structural objects. An erroneous restoration of details can induce errors in object detection and identification. Consequently, the authors underscore the significance of preserving context in tasks that necessitate scene interpretation, such as video surveillance and dashboard cameras. The yardstick and subjective comparison brought to light that different SR models perform divergently based on the types of degradation in the input and the motion of the camera. Some models exhibited overfitting to specific down sampling methods, while others demonstrated greater stability and yielded similar outcomes in different tests. The introduction of noise to the images typically diminished the metric values and altered the rankings of the models. The authors put forth the ERQAv2.0 metric, which outshines other quality metrics in terms of its correlation with subjective scores for detail restoration. This metric takes into account how well the structure of objects is restored relative to the ground-truth imagery. This study offers profound insights into the capabilities of SR models in restoring intricate details and accentuates the importance of preserving context. The yardstick and objective assessment metric can prove to be invaluable tools for future research in the realm of SR-based work.

In [6], the author invokes an interesting concept, claiming that the SRGAN is an innovative method for obtaining high resolution images from low resolution inputs. However, it is worth noting that SRGAN often results in images marred by unsightly imperfections. In this article, the authors present an Advanced SRGAN that seeks to improve the visual quality by optimizing three fundamental elements of SRGAN: network architecture, adversary loss, and perception loss. They introduce a novel building unit, the Residual in Residual Density Block, as the basis of the network. The authors choose to dispense with the use of batch normalization layers and instead utilize residual scaling and reduced initializations to facilitate the development of a deep neural network. Additionally, the authors further refine their discriminator by introducing the RaGAN structure., which assesses the relative authenticity of images. Additionally, they bolster the perceptual loss by utilizing features before activation, thereby providing more robust guidance for maintaining brightness consistency and recovering texture. The proposed ESRGAN consistently outperforms SRGAN in terms of visual quality, offering more lifelike and organic textures. It is noteworthy that it achieved the highest ranking in the Perceptual Index 2018-SR Challenge. The authors also provide the ESRGAN code. This article highlights the importance of Single Image Super-resolution  which is a process of reconstructing high-resolution image data from low-resolution data. SISR has captivated the attention of both the research community and AI companies. Although approaches focused on PSNR have made strides in improving the numerical value, they often yield overly smoothed outcomes lacking in high-frequency details. To shed light on the topic, the authors draw comparisons between ESRGAN and state-of-the-art PSNR-oriented methodologies such as SRCNN, EDSR, and RCAN, as well as perceptual-driven approaches like SRGAN and Enhance Net. Their findings reveal that ESRGAN surpasses previous techniques in terms of sharpness and detail, resulting in more authentic textures and fewer imperfections. In essence, the authors propose ESRGAN as an enhanced rendition of SRGAN, one that achieves superior visual quality through the enhancement of network architecture, adversarial loss, and perceptual loss. ESRGAN's performance eclipses that of prior methods, securing the top spot in a super-resolution challenge within the realm of SR-based endeavors.

As proposed in the paper [7],The MDCN is a novel deep learning approach for SISR. This innovative network incorporates both residual and dense connections to enhance the flow of information and address the issue of gradient vanishing. To further enhance its capabilities, the authors introduce a scale recurrent framework that allows the network to super-resolve images at different scales using the same set of convolutional filters.Comparing the MDCN to existing super-resolution techniques, it can be observed that it provides the most advanced performance for smaller scale factors (2 \& 3) and the most competitive performance for larger scale factors (4 \& 8). The training procedure includes a combination of low-level (L1) loss, high-level (VGG) loss, and high-level (Adversarial) loss to enhance both the objective (pixel-reconstructruction error) and visual (sharpness) quality of the super-resolution images.The paper delves into several aspects of the MDCN, including the impact of the scale recurrent design, variable depth supervision, number of dual-link units, order of addition and concatenation connections, and convergence curve. The findings reveal that the scale recurrent design significantly enhances performance, and the MDCN network proves to be efficient in terms of parameter size and convergence.Extending the MDCN approach to video super-resolution, the authors employ a multi-image restoration technique. This allows the network to process multiple neighboring low-resolution frames and achieve superior results compared to existing video super-resolution methods. Therefore, the proposed MDCN network not only excels in SISR but also exhibits potential in video super-resolution by leveraging the power of multiple frames.By incorporating both objective and perceptual loss functions, the MDCN network succeeds in enhancing the quality of the super-resolved images. The various aspects explored in the paper, such as the scale recurrent design and variable depth supervision, contribute to the efficiency and effectiveness of the MDCN network.

The paper[8] examines the concept of "image super-resolution", a technique that seeks to improve the resolution of low resolution images or videos. Deep learning-based techniques employed in super-resolution have resulted in significant improvements in image quality.. The paper introduces both supervised and unsupervised methods for super-resolution and explores the various convolutional neural networks (CNNs) employed in these approaches.Supervised super-resolution methods rely on paired low-resolution and high-resolution images for training. These methods can be further categorized into CNN-based techniques and those based on generative adversarial networks (GANs). CNN-based methods, such as SRCNN and LapSRN, utilize convolutional neural networks to acquire the knowledge of transforming low-resolution images into high-resolution counterparts.In contrast, GAN methods, such as SRGAN and ESGRAN, utilize the combined capabilities of both the generator and the discriminator to improve the perceived quality of the reconstructed images.Unsupervised super-resolution methods, in contrast, do not necessitate paired training data and have the ability to directly reconstruct low-resolution images from the real world. One such method is ZSSR, which trains a small image-specific CNN during the testing phase using solely the low-resolution image itself. Another method, IKC, employs an iterative correction process to estimate the blur kernel responsible for image degradation and subsequently refine the super-resolution outcome.The paper also delves into various challenges and future prospects in super-resolution research. These encompass the development of stable non-reference quality evaluation indicators, exploration of unsupervised super-resolution techniques, and the application of super-resolution models in real-world domains such as face recognition and medical imaging. In essence, the paper offers an overview of deep learning-based image super-resolution methods while emphasizing the disparities between supervised and unsupervised approaches. Furthermore, it delves into the specifics of CNNs utilized in these methods and provides valuable insights into the future directions of research in the field of super-resolution.

In [9], the authors present a novel approach to enhance the quality of images and videos by utilizing a recurrent generative adversarial network named SR2 GAN. To achieve remarkable precision in super-resolution tasks, they employ a convolutional neural network integrated with residual learning models.The proposed model employs a recursive approach to learn the transformation function necessary to synthesize high-resolution images, which is an effective solution to the problem of video super-resolution. This approach not only reduces model parameters and depth, but also guarantees the successful synthesis of high-quality images. The authors conduct comprehensive testing of the proposed model against existing methods and find that the SR2 GAN model outperforms them with respect to PSNR and SSIM. Additionally, the authors generously provide the source code and supplementary materials for the proposed model. The network architecture comprises an encoder-decoder structure featuring residual and inverse residual blocks. While the encoder captures an image's context through downsampling, the decoder reconstructs the image using the extracted context information. To maintain gradient flow and temporal consistency in video frames, skip connections and concatenation are employed. The model is meticulously trained on various datasets for image and video super-resolution, including renowned benchmark datasets like Set5, Set14, BSD100, Urban100, Apollos capes, and Cityscapes. The results demonstrate unambiguously that SR2 GAN produces high resolution images with improved visual sharpness and structural similarity in comparison to alternative techniques. The proposed approach is widely accepted as a deep, concise, and optimal solution for super-resolution image and video applications. STRB in India generously supported this work.

The [10] article introduces a novel framework called FREGAN (Frame Rate Enhancement Generative Adversarial Network), which aims to enhance the frame rate in videos. By leveraging a sequence of past frames, the model predicts future frames to effectively increase the frame rate. The researchers employed the Huber loss as a loss function in the FREGAN model, leading to exceptional results in super-resolution. To evaluate the model's performance, standard datasets such as UCF101 and RFree500 were utilized. The results demonstrated that the proposed model achieved an impressive PSNR of 34.94 and a SSIM of 0.95. These metrics were employed to assess the model's effectiveness. In this article, we provide a detailed description of a FREGAN model. The FREGAN model consists of a generator (generator) and a discriminator (discriminator). The generator predicts an intermediate frame on the basis of two adjacent frames. The discriminator network is used to distinguish between real and false frames. In the model, a customized Convolutional Neural Network (CNN) model is employed in the discriminator, while a modified version of the UNet architecture is used in the generator. The proposed model was trained on datasets comprising 256x256 pixel, 3-channel images extracted from 30fps and 24fps videos. The training process involved the use of the Adam optimizer, and the delta parameter of the Huber loss function was fine-tuned to achieve optimal results. The experiments revealed that an optimum delta value of 0.5 yielded the highest PSNR and SSIM scores. When compared to other methods for frame rate enhancement, the proposed model outperformed them in terms of SSIM scores and achieved a PSNR score close to the best performing method. The FREGAN model showcased robust performance, and the Huber loss function effectively handled noise in the images. This article introduces a promising approach to enhance the frame rate in videos, particularly in applications such as gaming and autonomous driving. Future advancements in the generator structure and the incorporation of additional intermediate frames have the potential to further enhance the model's performance.

the article [11] discusses the wide-ranging applications of super-resolution reconstruction in various fields such as medical imaging, security monitoring, and remote sensing imaging. This research primarily focuses on enhancing low-resolution images to generate high-resolution counterparts using a technique called SISR. The advancement of CNNs has greatly improved super-resolution techniques, and attention models have recently been introduced to filter unnecessary visual information and emphasize relevant features. This paper introduces a unique approach by combining a SRGAN with a CBAM to create a more lightweight network called (attention-model) AM-SRGAN. This integration is a significant contribution as it directly combines the attention module with GAN, offering a fresh perspective on optimizing GAN network structures. The CBAM-embedded SRGAN simplifies the network structure and enhances its representational capabilities, potentially reducing the need for excessively deep networks, which can be challenging to train. The paper explores the use of attention models in the context of GANs for super-resolution reconstruction. The combination of CBAM with SRGAN introduces a more efficient and lightweight network architecture, demonstrating the potential to simplify complex super-resolution techniques while maintaining or even improving image quality. The introduction of neural networks into super-resolution reconstruction has significantly improved its effectiveness. Researchers continuously strive for better results by exploring new network structures and frameworks. However, as networks become deeper, training time increases, and network robustness may decrease. Integrating attention models like the CBAM module into SRGAN offers an innovative approach. It filters important features, streamlining the generator's structure and accelerating critical reconstruction feature extraction. The experimental results confirm that the attention model optimizes the network's construction, reducing complexity and reconstruction time. This study provides valuable insights for GAN-based computer vision and highlights the untapped potential of attention models in super-resolution reconstruction, promising further research in this area

In our research paper [12], we utilize GANs as a framework for the generation of natural frames with remarkable perceptivity and realism. The GAN framework directs the generation process to regions in the search area that are more probable to contain photorealist frames, thus moving us closer to the domain of natural frames.. To accomplish this, they utilize a highly sophisticated Residual Network (ResNet) architecture within the GAN framework, setting a new benchmark in video super-resolution, particularly for high upscaling factors, as measured by PSNR. Our contribution is the introduction of a generator network that optimizes a unique perceptual loss, calculated using feature maps from the VGG network. This approach offers greater resilience to changes in pixel space compared to the traditional Mean Squared Error (MSE)-based content loss In the initial stage, our proposed algorithm was evaluated against various state-of-the-art video super-resolution methods. Following that, a qualitative assessment was conducted to evaluate the performance and outcomes of these different video super-resolution techniques. In order to conduct our experiments, we utilized a video database that is accessible to the public and contains high-quality material. This video series was utilized to train our algorithm and to conduct experiments. The resulting video demonstrates super-resolved frames generated using different video super-resolution methods.they presented a video super-resolution (SR) algorithm that utilizes a generative adversarial network (GAN). Our approach achieves a new benchmark performance, as measured by the widely recognized PSNR measure.We conducted a thorough analysis of different architectures and compared them to our algorithm. Our results demonstrate that GAN based improvements for significant scaling factors yield significantly more photorealistic reconstructions than those obtained from conventional reference methods.

The author in [13] explores the advancements of deep convolutional neural networks (CNNs) in video super-resolution (SR). However, the conventional approaches heavily rely on optical flow estimation for frame alignment, which often leads to the presence of artifacts. To overcome this challenge, we propose a novel end-to-end CNN that dynamically generates spatially adaptive filters using local spatio-temporal pixel channels. This innovative approach eliminates the need for explicit motion compensation, resulting in improved alignment and temporal consistency. Furthermore, we incorporate residual modules with channel attention to enhance feature extraction. Through comprehensive evaluations on three public video datasets, our method demonstrates superior performance in terms of image clarity and texture preservation compared to state-of-the-art techniques. While deep CNNs have made significant progress in video super-resolution, the dependence on optical flow for frame alignment poses difficulties and introduces artifacts. To address this, we introduce a comprehensive CNN that generates adaptive filters based on local spatio-temporal pixel channels, eliminating the requirement for explicit motion compensation. This approach enhances alignment and temporal consistency and facilitates high-resolution frame restoration in the reconstruction network. By incorporating residual modules with channel attention, we further improve feature extraction. Our approach out-performs existing techniques in terms of image clarity and
texture preservation, as evidenced by evaluations conducted on three public video datasets. The authors present a one-stage framework for Video Super-Resolution (VSR) that aims to reconstruct high-resolution videos without the need for explicit motion compensation. The authors suggest the implementation of STAN to achieve temporal convergence in the feature domain. This would improve the integration of multi-frame data and enhance the overall temporal consistency of the video. Furthermore, they introduce a comprehensive VSR network that is composed of the STAN for alignment and the reconstruction network for temporal information aggregation. This design enables the network to effectively analyze interframe spatial information and learn to synchronize neighboring video frames. Through extensive testing, it has been demonstrated that the STAN is more efficient than existing two-layer networks and can compete with the most advanced SISR and VSR techniques.

In [14], The author examines the need to improve the visual perception of compressed video, particularly in the context of the video coding standard H.265 /HEVC. While HEVC provides efficient compression, it unavoidably results in the loss of video information during compression and transmission, thereby impacting both objective and subjective quality. To tackle this issue, generative adversarial networks (GANs) have garnered significant attention for their capacity to reduce artifacts and elevate visual quality in image and video enhancement tasks. GANs accomplish this by acquiring the knowledge of image distribution mapping, thereby minimizing the disparity between generated and training set image distributions through adversarial loss. The evaluation of image and video restoration typically depends on full-reference techniques employing metrics like PSNR and SSIM. However, real-world transmission tasks often lack access to original images, necessitating the employment of non-reference assessments. The "Perceptual Index" (PI), which was introduced in the 2018 "PIRM Challenge," quantifies the difference between the image representation of the training dataset and the representation of the reconstructed video frames for the purpose of assessing visual quality without reference image. General-purpose annular systems (GANs) provide a basic principle for determining the ratio of objective distortion to visual perception quality through their loss functions. These functions include control distortion, which is the difference between the pixels between the actual image and the reconstructed image, as well as control perception quality, which is the ratio between the generated image and the natural image distribution. The paper outlines a wide range of image restoration techniques, including techniques that focus on resolving super-resolution images and reducing compression artifacts. Numerous methods harness the power of GANs, such as SRGAN and ESRGAN, which employ perceptual loss functions and adversarial training to enhance image quality. Drawing inspiration from the success of GANs in improving the visual perception of super-resolution images, This paper proposes the application of GAN properties to improve the visual performance of HEVC- compressed videos. It outlines a GAN-driven algorithm designed to improve the visual fidelity of HEVC compressed video. By reducing the disparity between the image distribution produced by the GAN generator (G) and the training set distribution via adversarial loss, it is possible to restore the visual fidelity of video frames. The HEVC-compressed images effectively direct the generation network of the GAN to learn the mapping from encoded images to their original images, a process that the GAN’s discriminator network continually improves upon. Our experiments demonstrate the effectiveness of our methodology.

In [15], the author explores the realm of artificial intelligence (AI), which has witnessed an extraordinary surge in growth and data accumulation. Within this realm, machine learning has emerged as a crucial player, particularly in its subset known as representation learning. This subset has proven to be invaluable in extracting valuable information from challenging data scenarios. One notable technique within representation learning is deep learning, which excels in deriving abstract, high-level features from data.Machine learning can be categorised into two distinct sub-categories: supervised and non-supervised.. The latter has gained increasing importance due to the difficulties associated with collecting labeled data. To address this challenge, generative models like Generative Adversarial Networks (GANs) have revolutionized unsupervised learning by harnessing game theory principles to generate realistic data, bypassing the need for Markov chains or approximate inference. The primary focus of this survey is to delve into the state-of-the-art in GANs, providing comprehensive insights into their definitions, motivations, applications, and training techniques. Furthermore, it explores various evaluation metrics and the diverse fields where GANs have found application. The survey concludes by analyzing the limitations of GANs and offering valuable insights into potential areas for future research in this dynamic and rapidly evolving field. In summary, this paper offers an extensive overview of the research context surrounding GANs, delving into their fundamental principles and exploring their derivative models and diverse applications across different domains. Additionally, it tackles the topics of evaluation metrics and training strategies. Finally, it highlights the current challenges faced by GANs and identifies potential avenues for future research.

In [16], The author proposes a novel concept that combines a model built on the outstanding GAN with an edge enhancement method for video super resolution reconstruction. The main goal of the model is to improve the image quality of low resolution and blurring videos, including those recorded by CCTV cameras. By employing the GAN framework, the model is able to generate remarkably realistic and intricately detailed super-resolution frames. This is achieved by training discriminators to differentiate between the enhanced frames and the original frames. Additionally, the model incorporates an edge enhancement function which further enhances the results by accentuating the edges through the utilization of the Laplacian edge module. Furthermore, the model integrates a perceptual loss mechanism to enhance the overall visual experience. The results of the experiments performed on a variety of datasets, including the well-known Vid4 dataset, demonstrate the robustness of the proposed approach. The article highlights the importance of video super resolution in computer vision and its broad-reaching applications in areas such as remote sensing or medical imaging. It examines the various types of super resolution, including SISR, and VSR., while also highlighting the advancements made possible through deep learning-based approaches. The article also acknowledges the challenges encountered when training deep learning models, particularly the need for copious amounts of paired LR-HR data and the complexities associated with the degradation process in real-world scenarios. In conclusion, the proposed model, which implements both GAN and edge enhancement techniques, exhibits highly promising outcomes in the domain of video super-resolution reconstruction. By combining the potential of deep learning with edge enhancement, this model has the ability to generate super-resolution frames that are not only lifelike but also immensely detailed. The conducted experiments successfully demonstrate the advantages of this proposed method across various datasets, including low-resolution videos acquired through CCTV cameras and other real-world scenarios. Ultimately, this model has the potential to significantly elevate the quality of low-resolution videos and enhance numerous computer vision tasks.

\section*{Conclusion}

The field of Video Super resolution and frame acceleration has had see an unprecedented amount of  research and studies due to the growing power of AI and its implication in generation a higher quality multimedia from old and poor quality video for better user experience. This survey of research papers highlights the diverse approaches and methodologies that various researchers have developed in order to solve the challenge of super resolution and frame generation. key takeaways from this exploration include the use of various SRGANs to generate the super resolute version of the frame and generate the details that might have been lost due to less resolution, the use of frame generation with the help of GANs by prediction the intermediate frame have been used to increase the smoothness of the video. The quality of the enhanced videos generated has been steadily improving. These innovations promise to restore damaged or substandard video and aid in the revival of cultural and scientific knowledge.

\vspace{12pt}
\end{document}